# Deconstructing the governing dissipative phenomena in the nanoscale

Sergio Santos[1,2], Carlo A. Amadei[1], Tzu Chieh Tang[1], Victor Barcons[2], Matteo Chiesa[1]

[1] Center for Future Energy Systems (*iFES*), Masdar Institute of Science and Technology, Abu Dhabi, 54224, UAE

[2] Departament de Disseny i Programació de Sistemes Electrònics**,** UPC - Universitat Politècnica de Catalunya Av. Bases, 61, 08242 Manresa, Spain

# Abstract

An expression describing the controlling parameters involved in short range nanoscale dissipation is proposed and supported by simulations and experimental findings. The expression is deconstructed into the geometrical, dynamic, chemical and mechanical properties of the system. In atomic force microscopy these are translated into 1) tip radius and tip-sample deformation, 2) resonant frequency and oscillation amplitude and 3) hysteretic and viscous dissipation. The latter are characteristic parameters defining the chemical and mechanical properties of the tip-sample system. Long range processes are also discussed and footprints are identified in experiments conducted on mica and silicon samples. The present methodology can be exploited to validate or invalidate nanoscale dissipative models by comparing predictions with experimental observables.

**I. Introduction**

Nanoscale dissipation should account for the complex interplay of phenomena related to the irreversible local rearrangement and displacement of atoms, atomic reorientation and relative motion between atoms and the energy losses involved in the formation and rupture of chemical and physical bonds[1-3]. Since in dynamic atomic force microscopy (dAFM) stored mechanical energy is transferred from a micro-cantilever to a sample's surface, in principle, dAFM provides the means to investigate such nanoscale phenomena and test the experimental validity of dissipative models[1, 4, 5]. The vibrating cantilever is well described in many cases of interest by a simple point mass model, a spring and a linear damper where non-linearities are introduced by the tip-sample force[6, 7]. A fraction of the energy stored in the driven micro-cantilever is irreversibly lost in the tip-sample interaction during every oscillation cycle via several dissipative nanoscale processes. It is the decoupling and understanding of these processes and their relevance in nanosciences that is attracting much attention in the community[8-16].

Here, an expression relating the controlling parameters involved in viscous and hysteretic processes during tip-sample mechanical contact is proposed and interpreted in the context of dAFM, and, in particular, in amplitude modulation (AM) AFM. The expression accounts for geometric, dynamic and material properties of the cantilever-tip-sample system and is based on the energy conservation principle and standard but simple and phenomenologically derived dissipative models employed in the literature[1, 17, 18]. Such expression has been recently reported[18] to predict how dissipation processes depend on the size of the interacting tip but the effects of the presence of simultaneous dissipative processes remained unclear. Here, both simulations and experimental tests confirm that the relative prevalence of a given short range dissipative process, i.e. viscous or hysteretic, can be tuned as predicted by both the geometrical and dynamic parameters that appear in the expression. The predictions are experimentally tested and verified on mica samples. The standard Sader-Jarvis-Katan formalism of force reconstruction[10, 14, 19] and energy expressions[12, 20] are further investigated in the presence of challenging long range hysteretic dissipative forces. These processes could account for long range capillary interactions or the onset of processes related to chemical affinity between the tip and the sample. Complimentary methodologies dealing with dissipative interactions are proposed, developed and verified via simulations and experiments. These complimentary methodologies are shown to overcome[1, 21] the standard shortcomings of conservative force reconstruction techniques. The predicted footprints of long range dissipation are experimentally tested and verified to occur on mica and silicon samples.

**II. Methods: the governing equation of motion, energy dissipation and force reconstruction**

**A. Reconstruction dissipative and conservative interactions in AM AFM**

The governing equation of motion in dAFM, is typically modeled with the use of a standard driven and linearly damped harmonic oscillator[22]

$$m\frac{d^2z}{dt^2} + \frac{m\omega}{Q}\frac{dz}{dt} + kz = F_{ts} + F_D \qquad (1)$$

where k is the spring constant, Q is the Q factor due to dissipation with the medium, $\omega$ is the angular drive frequency, the effective mass is $m=k/\omega^2$ and $F_D$ is the driving force. Typically the drive frequency is set equal to the natural frequency $\omega_0$ since this leads to convenient simplifications[12]. Furthermore, z is the instantaneous position of the tip relative to its unperturbed equilibrium position and $z_c$ is the tip-sample separation [22]. A geometrical constraint relates the two since the instantaneous tip-sample distance d can be written as $d=z_c+z$.

From (1), the energy dissipated per cycle $E_{dis}$ can be derived without imposing any restrictions in the nature of the dissipative force[12]

$$E_{dis}(d) = \frac{\pi k A_0 A(d)}{Q}\left[\sin(\Phi(d)) - \frac{A(d)}{A_0}\right] \quad (2)$$

where $A_0$ is the free (unperturbed) amplitude of oscillation, A is the perturbed oscillation amplitude and $\Phi$ is the phase lag. Note that $\Phi$ as obtained in an experiment could be termed $\Phi_{dis}$ and accounts[23] for dissipative and conservative contributions[2]; $\Phi$ and $\Phi_{dis}$ are used interchangeably here. The condition $E_{dis}=0$ in (2) allows defining a phase shift (conservative) $\Phi_c$[12]

$$\Phi_c(d) = \sin^{-1}\left(\frac{A(d)}{A_0}\right) \quad (3)$$

The absolute of the difference $\Delta\Phi(d)=\Phi_{dis}(d)-\Phi_c(d)$ (phase difference) can be employed to decouple[17, 21] hysteretic from viscous dissipative processes and it is employed in this work throughout. The Sader-Jarvis-Katan (conservative force) in terms of the minimum distance of approach $d_m$, where $d_m$ can be directly identified with d, reads[10, 19]

$$F_{ts}(d_m) = 2k\int_{u=d_m}^{u=\infty}\left[\left(1+\frac{A^{1/2}(u)}{8\sqrt{\pi(u-d_m)}}\right)\Omega(u) - \frac{A^{3/2}(u)}{\sqrt{2(u-d_m)}}\frac{d\Omega(u)}{du}\right]du \quad (4)$$

where $\Omega(d_m)$ is the normalized frequency shift

$$\Omega(d_m) = \left[1+\frac{A_0}{QA}\cos(\Phi(d_m))\right]^{1/2} - 1 \quad (5)$$

Higher modes are neglected by employing (1) alone and higher harmonics are further neglected in the derivation of (2)-(4)[7, 24]. The reader should note that neglecting the higher modes and higher harmonics might lead to significant deviations from the conclusions in this work only in highly damped mediums[25] where Q<10. For this reason, small amplitudes are sometimes employed in the literature to inhibit the excitation of higher modes[25, 26]. In ambient conditions however[6] $Q\sim10^2$-$10^3$ implying that the approximations are valid here. For example, the approximation $d_m=z_c-A$ leads to errors in $d_m$ in the order of 10 pm when employing cantilever properties similar to OLYMPUS AC160TS cantilevers ($f_0\approx300$ kHz, $k\approx40$ N/m and $Q\approx500$) and standard operational parameters as in this work (see supplementary Figure S1)[7, 25]. In the experiments in this work, a Cypher AFM from Asylum Research has been employed and the tip has been positioned at approximately 50 nm above the surface when using the thermal method[27] to calibrate $f_0$ ($\omega_0=2\pi f_0$) and Q.

In AM AFM care must be taken when monitoring d in amplitude and phase distance APD curves[28-30]. For example there are regions of the operational parameter space that do not allow recovering the range of d continuously. This is typically a consequence of bi-stability as described in the literature by several groups[29-32]. Also note that $A_0$ should be sufficiently large to establish mechanical contact with the surface in order to investigate short range dissipation in the mechanical contact region[22]. Since the recovery of $E_{ts}$ (d) and $F_{ts}$ (d) in this work depends on monitoring variations in d smoothly, a method to avoive bi-stability and reach regions where smooth force transitions occur is discussed in detail next; the reader can also refer to several works in the literature where the parameter $A_0$ is increased until force transitions occur smoothly and bi-stability is avoided[30, 33, 34].

1. First, for sufficiently small values of $A_0$, i.e. typically $A_0$<2-5 nm, the attractive regime prevails and mechanical contact is never established[35].

2. Increasing $A_0$ results in a range of $A_0$ values for which discrete transitions from the attractive force regime, where the phase lag Φ lies above 90º, to the repulsive force regime, where the phase lag lies below 90º, occur. The mean value of $A_0$ for which discrete transitions occur is termed critical amplitude or $A_c$[36]. Information about d is lost due to force transitions in this range.

3. Finally when, approximately[30, 35], $A_0$>2$A_c$, force, amplitude and phase transitions occur smoothly in d and the whole range of d is recovered from the APD curves (see supplementary Figures S1 and S2). In this work $A_0$>2$A_c$ throughout.

4. The concept of $A_c$ can further be employed to characterize the tip radius R in situ. This method has been used in this work throughout to monitor the tip radius[35]. According to the $A_c$ method, and for the AC160TS (OLYMPUS) cantilevers employed in the experiments in this work, R≈4.75$A_c^{1.12}$ (standard units should be employed).

5. The reader should note that force models in AFM typically assume an spherically terminated tip. Such models are also employed here. Experimentally this might not be a good approximation in some cases. Nevertheless, in the experiments in this work, an increase in the force of adhesion has been observed throughout with increasing relative humidity for both mica and silicon samples (data not shown). According to recent reports, this behavior is characteristic of tips that approximately terminate as spherically[37]. The authors acknowledge however that the presence of asperities[38] might lead to deviations from the results presented in this work[39].

Finally, here d has been set to zero in the expression d=$z_c$-A where minima in the tip-sample force $F_{ts}$ (d), i.e. the adhesion $F_{AD}$, occurs according to (5).

**B. Tip-sample force models**

In this section the conservative force models, i.e. long and short range, and dissipative, i.e. short range viscosity and hysteresis, and the numerical algorithms employed in the simulations when solving (1) are discussed. From the dissipative force models, an expression describing the controlling parameters involved in short range dissipation is derived and deconstructed into the geometrical, dynamic and mechanical properties of the tip-sample system.

In the long range, i.e. d>$a_0$, conservative London dispersion forces have been considered here following Hamaker's approach[40]

$$F_{vdW} = -\frac{HR}{6d^2} \qquad d>a_0 \qquad (5)$$

where vdW stands for van der Waals and H is the Hamaker constant. The term $a_0$ is an intermolecular distance; here $a_0$=0.165 nm throughout[41]. When mechanical contact occurs, i.e. d ≤$a_0$, the conservative force has been modeled with the standard Derjaguin-Muller-Toporov (DMT) model of contact mechanics[42]

$$F_{DMT}(d) = -\frac{HR}{6a_0^2} + \frac{4}{3}E^*\sqrt{R}\delta^{3/2} \qquad d \leq a_0 \qquad (6)$$

where $E^*$ is the effective elastic modulus of the contacting bodies and δ is the tip-sample deformation, i.e. $\delta = a_0 - d$[43, 44].

Viscosity, or velocity dependent dissipation occurring in the contact region, can be modeled with a Kelvin-Voigt model[1]

$$F_\eta = -\eta(R\delta)^{1/2}\dot{\delta} \qquad d \leq a_0 \qquad (7)$$

where $(R\delta)^{1/2}$ is the contact radius according to the DMT[1] model, $\dot{\delta}$ is the velocity of deformation and η is the tip-sample's characteristic viscosity. Integration of (7) over a cycle[44] gives the energy dissipated via viscosity $E_\eta$

$$E_\eta = \frac{\sqrt{2}}{4}\pi R^{1/2}\omega\eta A^{1/2}\delta_M^2 \qquad (8)$$

where $\delta_M$ is the maximum tip-sample deformation per cycle[17, 44].

Hysteretic dissipation in the contact region, i.e. d>$a_0$, can be modeled by assuming that the surface energy varies on tip approach and retraction [1, 2]

$$F_\alpha = -4\pi R\gamma\alpha = -4\pi R\Delta\gamma \qquad d \leq a_0 \text{ and } \dot{\delta} \geq 0 \qquad (9)$$

where $\gamma\alpha = \Delta\gamma$ is an increment in surface energy γ on tip retraction[2, 45]. Integration of (9) over a cycle gives

$$E_\alpha = 4\pi R\Delta\gamma\delta_M \qquad (10)$$

If (8) and (10) account for most of the dissipation in the interaction, the ratio $E_R = E_\eta/E_\alpha$ provides information about the controlling parameters in nanoscale dissipation as[17]

$$E_R = \sqrt{2}\,\frac{\delta_M}{R^{1/2}}\left[\frac{\eta}{\Delta\gamma}\right]\omega A^{1/2} \tag{11}$$

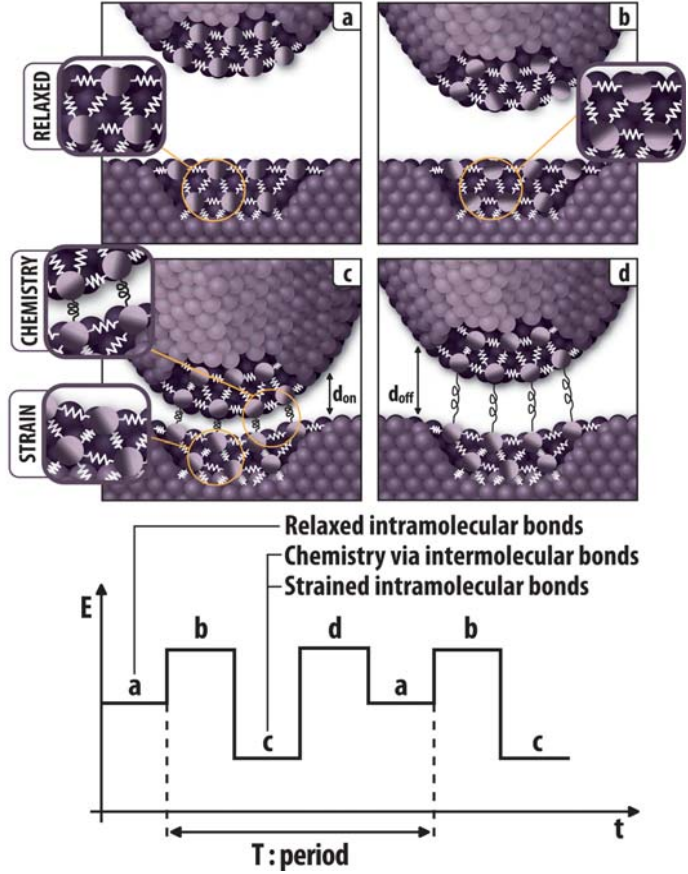

Figure 1. a-d) Illustrations representing electron and spatial configuration of atoms in a tip and a sample. The scheme describes the phenomena that might be involved during induced mutual tip-sample intermolecular and intermolecular interactions as the tip approaches and retracts from the sample. A scheme illustrating the evolution of the interaction energy E during a full oscillation period is shown at the bottom.

From (11), short range hysteretic and viscous forces will compete to control dissipation via η, Δγ, $\delta_M$, R, ω and A. The term in brackets contains the mechanical η and chemical Δγ material properties, the geometry of the system controls dissipation via $\delta_M$ and R and the dynamics of the system control dissipation via ω, A and $\delta_M$.

Qualitatively, and from (11), the prediction is that the larger $\Delta\gamma$ and R the larger the hysteretic contribution. This could be interpreted by a greater number of atoms involved in chemical interactions via $\Delta\gamma$ with increasing R. This phenomenon is illustrated in Figure 1 where, as the tip is sufficiently close to the surface (Figures 1c and 1d, intermolecular bonds are induced to decrease the energy E of tip-sample system. In the illustration also shows the bonds form at a distance $d_{on}$ (Figure 1c) on tip approach and rupture at a distance $d_{off}$ (Figure 1d) on tip retraction. Also from (11), the larger $\eta$, $\omega$, A and $\delta_M$ the larger the energy dissipated via viscosity. The interpretation is that the parameters $\omega$, A and $\delta_M$ are related to viscosity in that they affect the tip velocity and velocity dependent dissipation, i.e. viscosity. An interpretation is given in the illustration in Figures 1c and 1d where, the larger the tip velocity and deformation, the larger the induced strain on intermolecular bonds leading to dissipation. Quantitatively, if $E_R$ is smaller than 1, hysteretic dissipation is dominant. On the other hand, values of $E_R$ larger than 1 define a viscosity dominant interaction. The numerical and experimental verification of the above predictions, in terms of short range interactions, forms the basis of this work. In particular, the hypothesis is that parameters that can be tuned by the user and/or that are independent of the sample's properties, i.e. $A_0$, A, $\omega$, R, and $\delta_M$, might be employed to control the dominating dissipative process. Long range dissipative processes are discussed separately in the next section. Finally, in the simulations (1) has been implemented in C and solved with the use of a standard fourth order Runge-Kutta Algorithm.

**C. Samples**

Mica and silicon Si samples have been employed in the experiments. The mica sample was 0.21mm thick (highest quality grade V1). The Si sample was an electronic grade (>99.9%) <111> P-type Si chip (10X10mm) of thickness 460-530μm. Both samples are purchased by Ted Pella, Inc.

## III. Results

### A. Numerical results: conservative forces and short-range dissipative processes

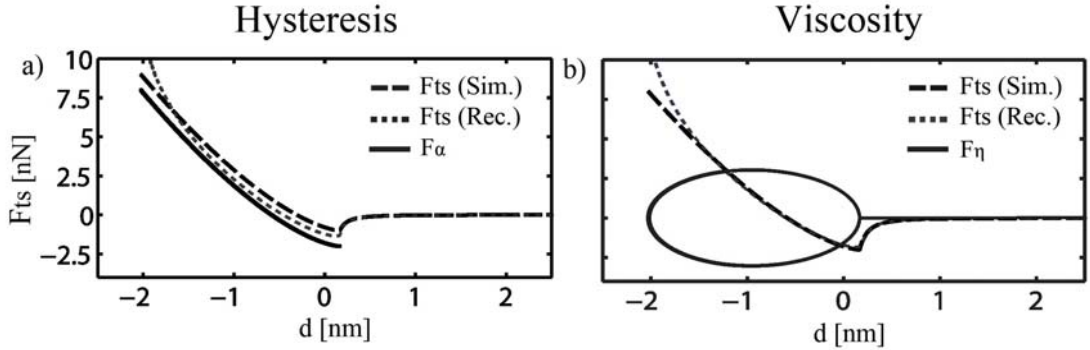

Figure 2. Tip-sample interaction profiles displaying conservative forces (long dashed) and where a) hysteresis (retraction path in continuous line) an b) viscosity (continuous line) in the short range are present. $F_{ts}$ (Sim.) stands for the conservative forces used in simulations (long dashed) and $F_{ts}$ (Rec.) (short dashed) is the recovered conservative force according to the Sadar-Jarvis-Katan formalism in (4).

In Figure 2 conservative forces (5)-(6), 1) hysteretic (9) (Figure 2a) and 2) viscous (7) (Figure 2b) short range dissipation have been implemented in the simulations. The figure illustrates how conservative forces can be recovered with the use of (4) in the presence of these short range dissipative forces. For Figure 2: k=40 N/m, $f_0$= 300 kHz, Q=450, $A_0 \approx$ 25 nm, $E_t$=120 GPa (Young modulus of the tip), $E_s$=1 GPa (Young modulus of the sample) and R=8 nm. Furthermore, in Figure 2a: γ=10 mJ, Δγ= 10 mJ and η=0 Pa·s (hysteretic dissipation only). In Figure 2b: γ=20 mJ, Δγ= 0 mJ and η=

100 Pa·s (viscous dissipation only). The conservative force is $F_{ts}$ (Sim.), where Sim. stands for simulation, and it is shown with long dashed lines. The recovered conservative force $F_{ts}$ (Rec.), where Rec. stands for recovered, according to (4) is shown with the use of short dashed lines.

In Figure 1a, the hysteretic force (9) acting only on tip retraction has been implemented and it is shown with the use of continuous lines. Two outcomes are worth noting. First, in the metastable region where (9) acts, $F_{ts}$ (Rec.) does not coincide with $F_{ts}$ (Sim.) but takes values in between the approach (conservative) and retraction (hysteresis) paths. The practical implication is that values recovered from (4) in the presence of metastability, or hysteretic forces such as those modeled by (9), i.e. surface energy hysteresis[4], should not be interpreted as originating from purely conservative phenomena. Second, as repulsive forces increase with decreasing d, $F_{ts}$ (Rec.) diverges from $F_{ts}$ (Sim.). This is a consequence of the small perturbation assumption when deriving (4)[14, 46] and implies that the approximation worsens with increasing indentation or deformation.

In Figure 2b short range viscosity (continuous lines) (7) has been introduced in the simulations. $F_{ts}$ (Rec.) coincides with $F_{ts}$ (Sim.) in this case with errors smaller than 5% in agreement with previous studies[19]. Still, as indentation, peak forces or deformation increase, and as in the case of Figure 2a, $F_{ts}$ (Rec.) eventually diverges from $F_{ts}$ (Sim.) due to the small perturbation assumptions. The match between $F_{ts}$ (Rec.) and $F_{ts}$ (Sim.) in Figure 2b is a consequence of the dissipative viscous force in (7) being an odd function of velocity[10]. Note that dissipative forces being an odd function of velocity is a fundamental assumption in the derivation of (4)[10]; for example, hysteretic forces are not odd functions of velocity, hence the mismatch between $F_{ts}$ (Sim.) and $F_{ts}$ (Rec.) in Figure 2a. In summary, this example shows that (4) applies with errors smaller than 5%, and provided peak repulsive forces are not too large, when the dissipative forces present in the interaction are odd functions of velocity.

Importantly, the results from Figure 2 indicate that (4) alone cannot be employed to identify or decouple short range hysteretic and viscous dissipation processes. Identifying and decoupling these processes via experimental observables however is required in order to experimentally verify the predictions of (11). Fortunately, alternative methods to identify dissipative processes have been proposed that make use of the energy dissipation $E_{dis}$ expression (2) and ingenuity[1, 21]. For example, Garcia et al. first proposed[1] that dissipative mechanisms could be identified in terms of the derivative of $E_{dis}$ with respect to oscillation amplitude A. In particular, the derivative is carried out in terms of normalized values $dE_{dis}^*/dA^*$ throughout where asterisks imply normalizing, i.e. $A^* = A/A_0$ and $E_{dis}^* = E_{dis}/E_{dis}(M)$ where $E_{dis}(M)$ is maxima in $E_{dis}$. Typically, $dE_{dis}^*/dA^*$ is plotted in the vertical axis with $A^*$ in the horizontal axis. Here, an alternative method is employed that simultaneously exploits $E_{dis}^*$ and the concept of (normalized) phase difference $\Delta\Phi^*$. Specifically the product $E_{dis}^*\Delta\Phi^*$ is employed here to identify short range viscous and hysteretic dissipation. Note that while this concept has already been applied[17, 21] in terms of $A^*$, i.e. $E_{dis}^*\Delta\Phi^*(A^*)$, here a further development is introduced by plotting $E_{dis}^*\Delta\Phi^*$ in terms of the more intuitive (normalized) tip-sample distance $d^*$, i.e. $E_{dis}^*\Delta\Phi^*(d^*)$. Normalization is carried out in terms of the absolute of the minima in d. The use of the signals a-b) $E_{dis}^*$, c-d) $\Delta\Phi^*$ and e-f) $E_{dis}^*\Delta\Phi^*$ as a function of $d^*$ (left column) and $A^*$ (right column) is demonstrated in Figure 3 where the forces in Figure 2 have been implemented to solve the equation of motion in (1).

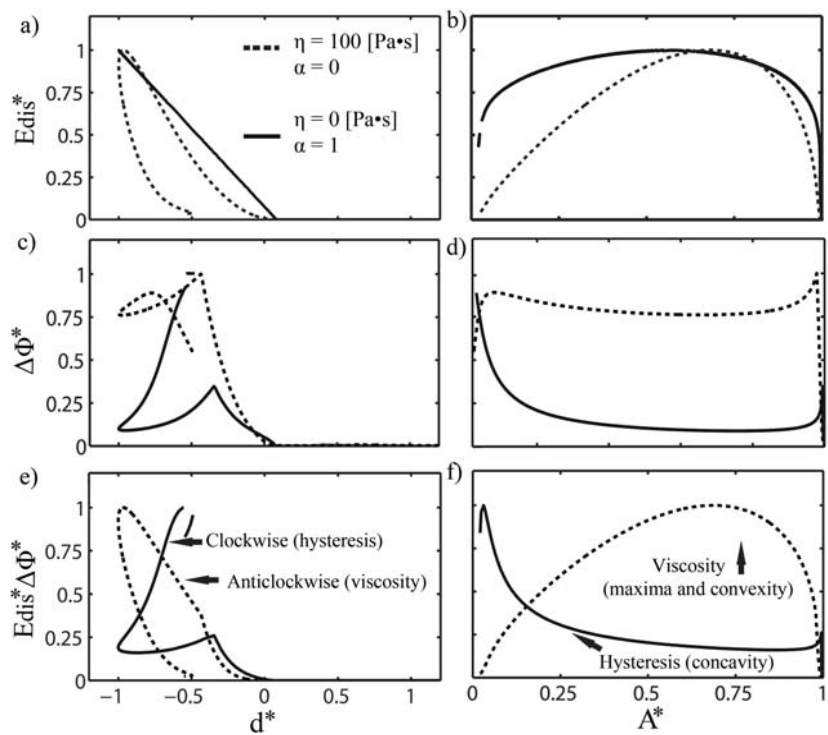

Figure 3. Simulations that explore the behavior of the a-b) $E_{dis}^*$, c-d) $\Delta\Phi^*$ and e-f) $E_{dis}^*\Delta\Phi^*$ signals in the presence of short range hysteresis $\alpha=1$ (continuous lines) and viscosity $\eta=100$ Pa·s (dashed lines) as a function of distance $d^*$ (left column) and amplitude $A^*$ (right column).

First note in Figure 3a that the $E_{dis}^*$ signal is two valued in $d^*$ when viscosity is present (dashed lines) $\eta>0$ and single valued when hysteresis is present $\alpha>0$ (continuous lines). Both signals are single valued in $A^*$ (Figure 3b). The corresponding phase difference $\Delta\Phi^*$ channel is shown for completeness in Figures 3c-d but it is not discussed in terms of the short range dissipation because it provides only limited information[21]. The $E_{dis}^*\Delta\Phi^*$ channel (Figures 3e-f) is more interesting here since it can provide information about the presence/absence of short range viscosity and hysteresis both in terms of d* and $A^*$. The main outcome is that when monitored in terms of d* (Figure 3e), and in the presence of short range hysteresis (continuous lines), the sense of rotation of the $E_{dis}^*\Delta\Phi^*$ signal is clockwise (always referring to decreasing $A^*$). In the presence of short range viscosity (dashed lines) the signal flips sense to anticlockwise. In terms of $A^*$ (Figure 3f) and in the presence of short range hysteresis (continuous grey line), concavity in $E_{dis}^*\Delta\Phi^*$ is observed at intermediate values of $A^*$. $E_{dis}^*\Delta\Phi^*$ also monotonously increases with decreasing $A^*$ except at the extreme values of $A^*$. In the presence of short range viscosity (dashed black line) convexity and maxima in $E_{dis}^*\Delta\Phi^*$ are observed at intermediate values of $A^*$. From the results in Figures 3e-f it can be concluded that the $E_{dis}^*\Delta\Phi^*$ signal is an experimental observable that can be employed to identify the presence of short range hysteresis or viscosity in terms of both d* (sense of rotation) and $A^*$ (behavior with decreasing $A^*$). The next question to address in terms of (11) however is whether the dominance of one or other mechanism, i.e. short range viscosity and hysteresis, can be deduced from $E_{dis}^*\Delta\Phi^*$ when both process are present in the interaction.

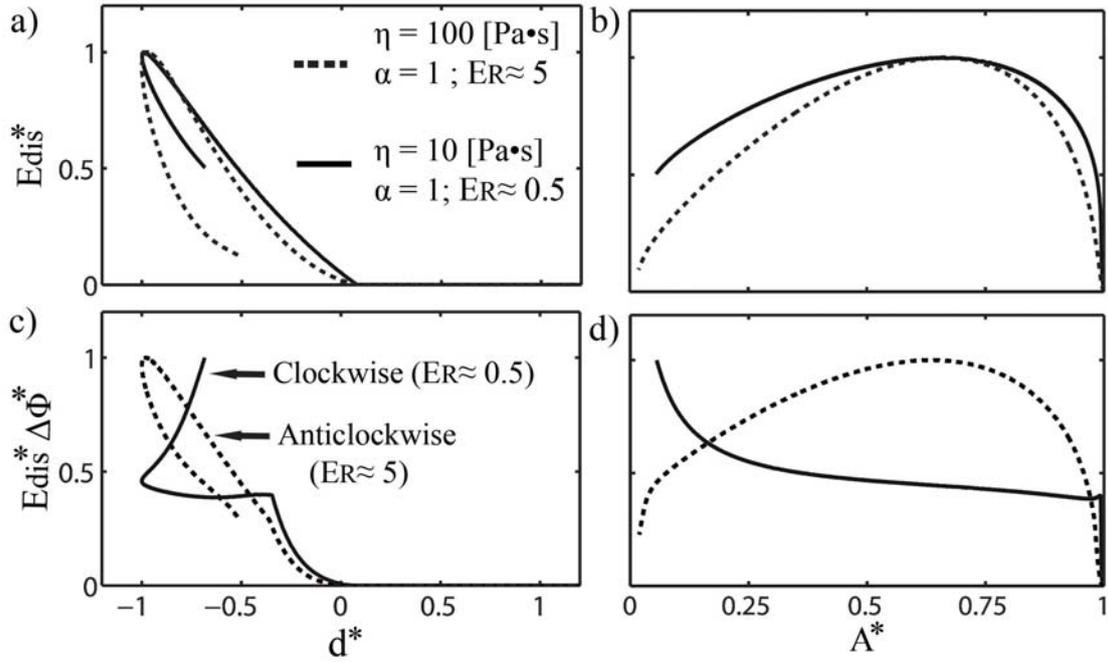

Figure 4. Simulations illustrating the different behavior of the a-b) $E_{dis}^*$, and c-d) $E_{dis}^*\Delta\Phi^*$ signals as a function of distance $d^*$ and amplitude $A^*$. Short range hysteresis $\alpha=1$ and viscosity $\eta>0$ are present in the interaction simultaneously. The behavior of the $E_{dis}^*\Delta\Phi^*$ signal presents characteristic footprints of both competing dissipative processes but the dominant process can still be identified, particularly in terms of $d^*$, by monitoring the sense of rotation of $E_{dis}^*\Delta\Phi^*$.

In Figure 4 short range viscosity (7) and hysteresis (9) have been considered in the simulations simultaneously. The conservative forces and numerical values are as in Figures 2 and 3. However, in Figure 4 one case involves $\alpha=1$ ($E_\alpha\approx14$ eV) and $\eta=100$ Pa·s ($E_\eta\approx70$ eV) (dashed lines) and the other case $\alpha=1$ ($E_\alpha\approx15$ eV) and $\eta=10$ Pa·s ($E_\eta\approx8$ eV) (continuous lines). The values of energy quoted represent maxima in eV for each process in each case. It follows that in the first case (dashed lines) $E_\eta>E_\alpha$ while in the second $E_\eta<E_\alpha$. The corresponding ratios $E_R=E_\eta/E_\alpha$ from (11) produce $E_R\approx5$ and $E_R\approx0.5$ respectively on the basis of differences in sample's properties alone. The behavior of the $E_{dis}^*$ signal is shown in Figures 4a-b. It is the $E_{dis}^*\Delta\Phi^*$ signal plotted against $d^*$ however that provides a powerful discrimination. In short, while $E_R$ cannot be computed directly in experiments, it follows that the anticlockwise rotation of the experimental observable $E_{dis}^*\Delta\Phi^*$ in Figure 4c (dashed lines) implies that $E_R>1$, or in other words that, $E_\eta>E_\alpha$. That is, short range viscosity is correctly predicted to be dominant over short range hysteresis. The inversion to clockwise rotation in the same figure (continuous lines) indicates that $E_R<0.5$. That is, again, $E_{dis}^*\Delta\Phi^*(d)$ correctly predicts that short range hysteresis dominates over short range viscosity. Similar predictions can be concluded from $E_{dis}^*\Delta\Phi^*(A^*)$ (Figures 4b and 4d), while, arguably less obvious. Figure 4 confirms the power of the $E_{dis}^*\Delta\Phi^*$ method to determine the relative dominance of short range dissipative processes. In summary, the $E_{dis}^*\Delta\Phi^*$ signal can be employed to validate/invalidate the predictions of (11) in simulations and experimentally. For example, similar simulations can be employed to show that the ratio in (11) is sensitive to geometric parameters such as the tip radius $R$ as predicted and that the sense of rotation of $E_{dis}^*\Delta\Phi^*$ agrees with this outcome (see supplementary Figure S3 for details).

## B. Experimental outcomes: conservative forces and short-range dissipative processes

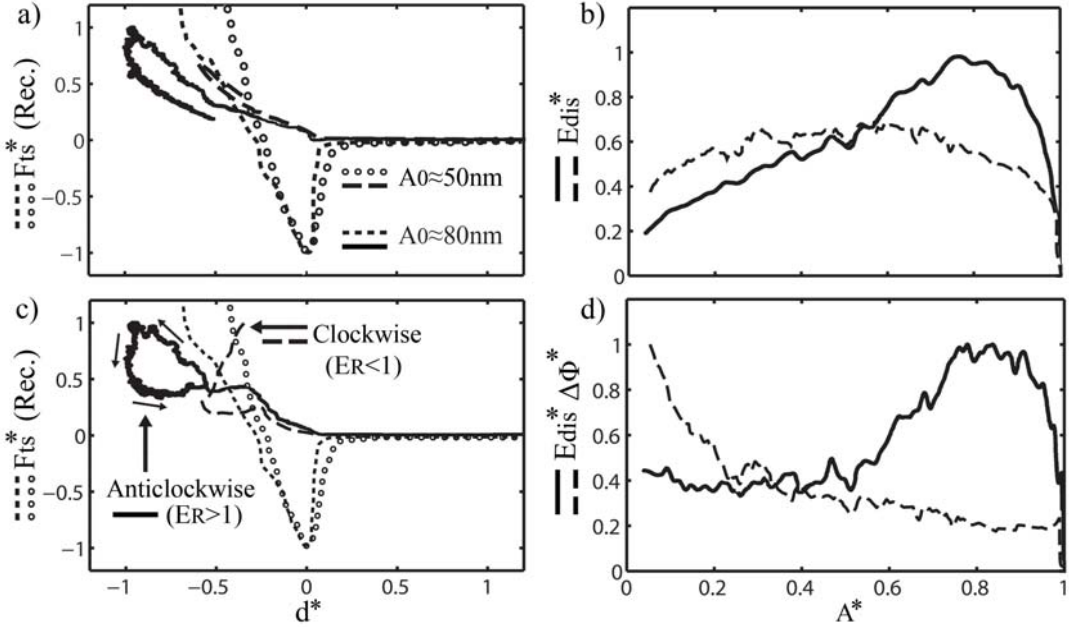

Figure 5. Experimental outcomes in terms of the normalized a, b) $E_{dis}^*$ and c,d) $E_{dis}^*\Delta\Phi^*$ signals as a function of distance a, c) $d^*$ and amplitude b, d) $A^*$ for a silicon tip tapping over a mica sample. The recovered normalized conservative force $F_{ts}^*$ is shown in a) and c) in short dashed lines and circles for $A_0\approx 50$ and 80 nm respectively. Variations in A and $A_0$ are investigated here. Normalization has been carried out with $F_{AD} \approx 6$ nN, $E_{dis} \approx 280$ eV, $E_{dis}\Delta\Phi \approx 0.58$ eV º ($A_0$=50nm) and 0.65 eV º ($A_0$=80nm) and $d \approx 5.6$ nm.

In Figure 5 a mica sample has been probed by varying $A_0$ to test the implications of variations in oscillation A and free $A_0$ amplitudes on tip-sample deformation δ (d<0), adhesion force $F_{AD}$ and on the $E_{dis}^*$ and $E_{dis}^*\Delta\Phi^*$ signals. According to the $A_c$ method[35] the tip radius in Figure 5 was R≈12 nm. In Figures 5a and 5c the normalized force $F_{ts}^*$(Rec.) is shown for $A_0$≈50 (circles) and $A_0$≈80 nm (short dashed lines). For both values of $A_0$ $F_{AD} \approx -6$ nN. The respective $E_{dis}^*(d^*)$ and $E_{dis}^*\Delta\Phi^*(d^*)$ signals are plotted in these figures, i.e. a and c. In Figures 5b and 5d the $E_{dis}^*(A^*)$ $E_{dis}^*\Delta\Phi^*(A^*)$ signals have also been plotted for both values of $A_0$.

First note from Figures 5a and c that maxima in both $E_{dis}$ and deformation $\delta$ increase from $E_{dis} \approx 190$ eV and $\delta \approx 3.8$ nm to $E_{dis} \approx 280$ eV and $\delta \approx 5.6$ nm respectively as a $A_0$ increases from 50 to 80 nm (in the figures $E_{dis}$ is normalized producing $E_{dis}^* \approx 0.7$ and 1 respectively). This increase of $E_{dis}$ and $\delta$ with $A_0$ is in agreement with both theory[1,2] and experiments[47,48]. It is the $E_{dis}^*\Delta\Phi^*$ channel however, both when monitored versus $d^*$ (Figure 5c) and $A^*$ (Figure 5d), that provides distinct information about the dominant dissipative interactions in each case. In particular, the sense in $E_{dis}^*\Delta\Phi^*$ varies (Figure 5c) from clockwise to anticlockwise as $A_0$ increases from 50 to 80 nm. From (11) and Figures 3e and 3c the change in the sense of rotation implies that both short range viscous and hysteretic forces are present in the interaction. Moreover, also form the definition of $E_R$ in (11), the change in sense of rotation implies that $E_R<1$ when $A_0 \approx 50$ nm and $E_R>1$ when $A_0 \approx 80$ nm. The practical implication is that one can control the dominant dissipative mechanism in the short range by simply decreasing (hysteretic dominant) or increasing (viscous dominant) the free amplitude $A_0$. Similar conclusions can be deduced by monitoring $E_{dis}^*\Delta\Phi^*(A^*)$ (Figure 5d) and by comparing it with Figures 3f and 4d. The physical phenomena leading to viscous-like (or velocity related) dissipation when a silicon tip interacts with a mica sample is not identified with the $E_{dis}^*\Delta\Phi^*$ channel in Figure 5 or model (7). Nevertheless, the results provide experimental evidence regarding the qualitative validity of (11), and the respective force models, i.e. (7) and (9), and predictions in terms of nanoscale dissipation. More thoroughly, the results in Figure (5) indicate the functional relations that the different dissipative nanoscale force models should satisfy with respect to the relevant parameters, for example A and $A_0$ in (7)-(11).

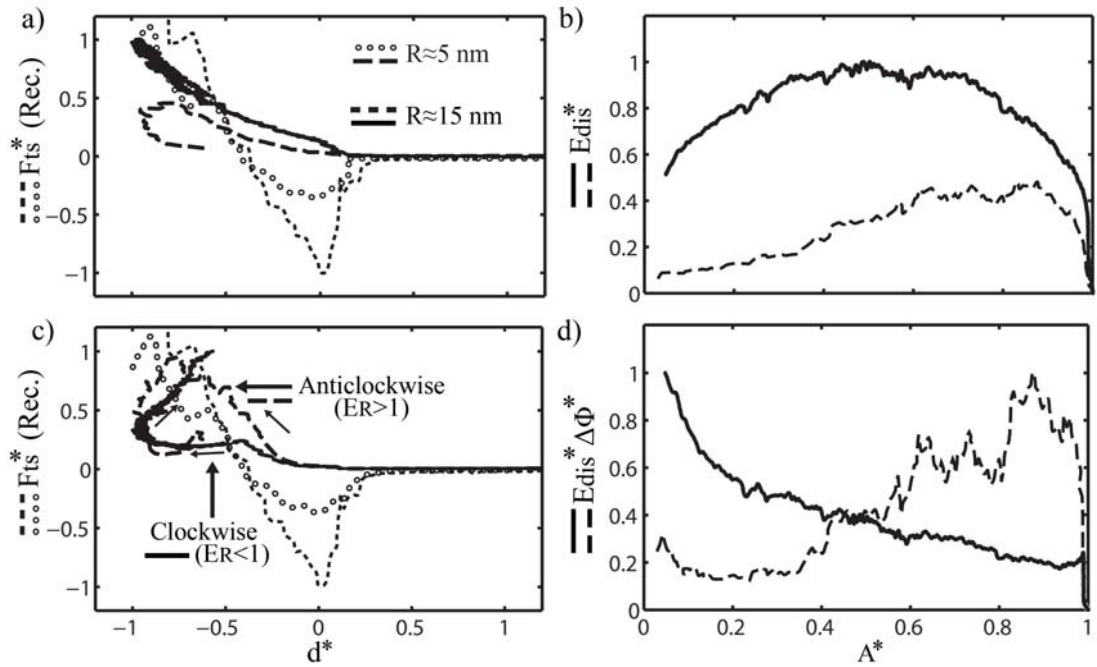

Figure 6. Experimental outcomes in terms of the a-b) $E_{dis}^*$, and c-d) $E_{dis}^*\Delta\Phi^*$ signals as a function of distance $d^*$ and amplitude $A^*$ for a silicon tip tapping over a mica sample. Variations in R are investigated here . The recovered conservative force $F_{ts}^*$ is shown in a) and c) as in Figure 5. Here, $A_0 \approx 50$ nm.

In Figure 6 a mica sample has been probed as a function of tip radius R with values $R \approx 5$ nm and $R \approx 15$ nm - R predicted by the $A_c$ method[35]. The free amplitude has been kept constant to $A_0 \approx 50$ nm. The tip-sample deformation was similar with both values of R, i.e. $\delta \approx 3.5$ nm, while the adhesion force approximately tripled, i.e. $F_{AD} \approx -2.4$ (circles) to -6.5 nN (short dashed lines) with increasing R. The energy signals $E_{dis}^*$ (Figures 6a-b) show that maxima in energy dissipation also approximately doubles from $E_{dis} \approx 105$ eV (long dashed lines) lines to $E_{dis} \approx 210$ eV (continuous lines) with increasing R. Again, the $E_{dis}^*\Delta\Phi^*$ channel provides information about transitions in dissipative phenomena both in terms of d* (Figure 6c) and A* (Figure 6d). The transition from anticlockwise (long dashed lines) to clockwise (continuous lines) in Figure 6c occurs by increasing R alone. Note also that $E_{dis}^*\Delta\Phi^*$ (A*) agrees with this interpretation (Figure 6d) as predicted from Figure 4d (see also Figure S3 in the supplementary material). From (11) and the $E_{dis}^*\Delta\Phi^*$ signals in the figure the conclusion is that the variation in R alone has resulted in a transition in the dominance of the dissipative process from viscous (R≈ 5 nm) to hysteretic (R≈ 5 nm). Again, as with the results in Figure 5, this conclusion is in agreement with the predictions of (11) and its interpretation.

Finally, a technicality should be mentioned with regards to the ratio $E_R$ proposed in (11). The ratio involves maxima in the energy curves corresponding to short range viscosity $E_\eta$ (8) and hysteresis $E_\alpha$ (10), i.e. $E_R = E_\eta/E_\alpha$. Nevertheless, the rotation of $E_{dis}^*\Delta\Phi^*$ occurs at the point of maxima in deformation $\delta$. Maxima in $\delta$ on the other hand might not coincide with maxima in the corresponding $E_\eta$ and $E_\alpha$ dissipative mechanisms (data not shown). This has direct consequences on the predictions discussed above in terms of the $E_{dis}^*\Delta\Phi^*$ channel and its ability to identify the dominant dissipative process. In particular, from this technicality, errors of approximately 10-20% in energy could follow in the predictions of the dominant short range process, i.e. viscosity or hysteresis.

## C. Long range dissipation and hysteretic phenomena

In dynamic AFM, long range dissipation has been typically related to hysteretic mechanisms that set on tip approach at a distance $d_{on}$ and break at a distance $d_{off}$[36, 49, 50] where $d_{on}>0$ and $d_{off}>d_{on}$[51] (see Figure 1). In ambient conditions $d_{on}$ could be identified with the distance at which the formation of a capillary neck occurs on tip approach and $d_{off}$ could be identified with the rupture of the neck on tip retraction[36]. In this section, this type of $d_{on}/d_{off}$ long range dissipative mechanism is investigated. The modeling is kept simple and assumes that dissipation occurs due to an additional (constant) hysteretic force $F_\alpha$

$$F_\alpha = -4\pi R \gamma \alpha_{nc} = -F_{AD}\alpha_{nc} \qquad (12)$$

where $F_\alpha$ acts when $d<d_{on}$ on tip approach and when $d>d_{off}$ on tip retraction.

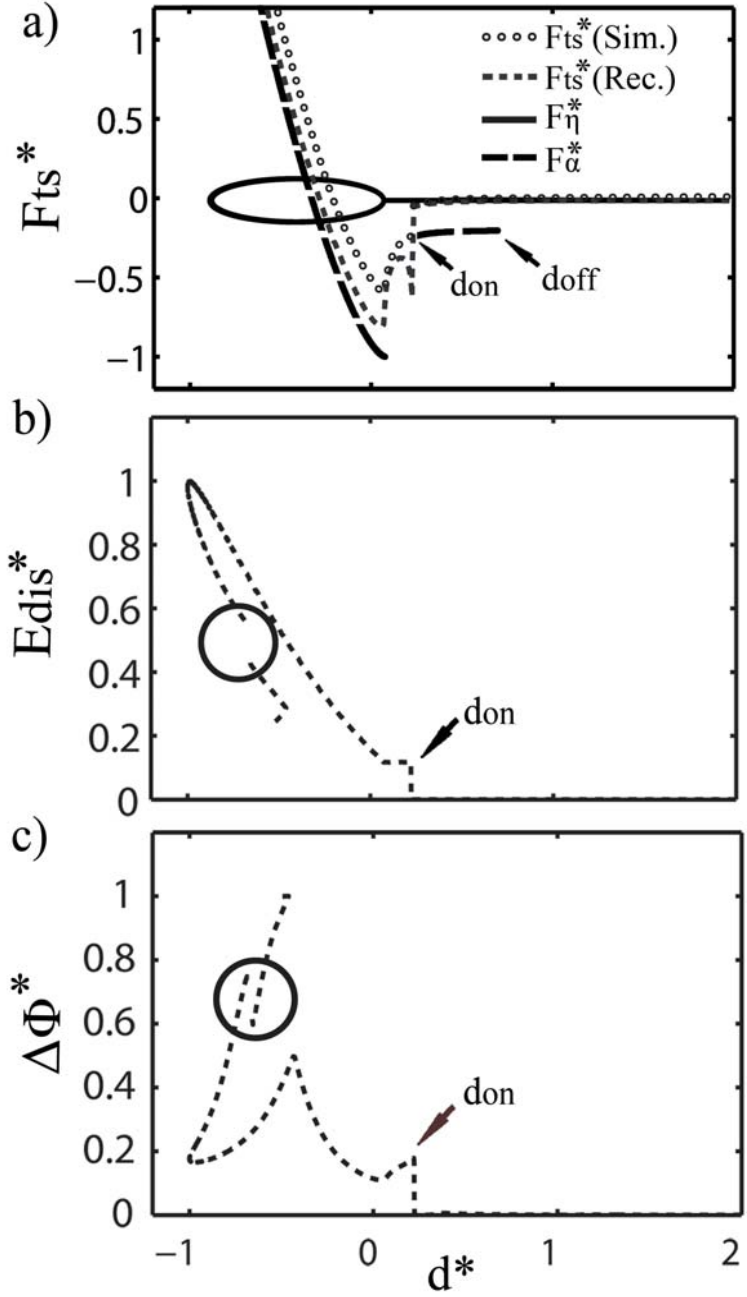

Figure 7. a) Tip sample (normalized) interaction displaying a conservative component $F_{ts}^*$ (Sim.) in circles, short range viscosity $F_\eta^*$ in continuous lines and short range hysteresis $F_\alpha^*$ in long dashed lines. A long range hysteretic component acting in the range $d_{on}<d<d_{off}$, is also present and also termed $F_\alpha^*$ (long dashed lines) where $d_{on}<d<d_{off}$. In this new range $d_{on}<d<d_{off}$ the force $F_{ts}^*$ (Rec.) (short dashed lines) only recovers the conservative force on tip approach and misses the path of tip retraction.

The corresponding b) $E_{dis}^*$ and c) $\Delta\Phi^*$ signals display a step at the distance $d=d_{on}$ and a second step at smaller distances $d^*$ and amplitudes $A^*$ (circles).

In Figure 7a a force profile displaying conservative and dissipative forces as in Figure 4, and implemented with similar parameters, is shown. In this case however, the long range hysteretic force (12) has also been implemented where $d_{on}$=0.5 nm, $d_{off}$=1.5 nm and $\alpha_{nc}$=0.5. Note also that $\alpha$=1 and $\eta$=10 Pa·s as in Figure 4 (continuous grey lines). Figure 7 is used here to discuss the simultaneous presence of long and short range dissipation, i.e. short range viscosity (7), short range hysteresis (9) and long range hysteresis (12). The normalizing parameters are: $F_{AD}$ ≈2.5 nN and d ≈2.3 nm for forces and distance respectively. The discussion of Figure 7a is similar to that of Figure 4 (continuous lines) and will not be repeated here except for the difference that (12) adds. In particular, $F_{ts}^*$ (Rec.) from (4) recovers (short dashed lines) the approach path of the force only in the range $d_{on}$<d<$d_{off}$. That is, information on the retraction path for $d_{on}$<d < $d_{off}$ is lost. Moreover, a drop occurs in $F_{ts}^*$ (Rec.) in Figure 7a at $d=d_{on}$. This distance corresponds to the onset of (12). In terms of affecting $F_{ts}^*$ (Rec.) (4), these are the main differences between dissipation occurring via hysteretic forces where $d_{on}$=$d_{off}$ as in (9) and hysteretic forces where the onset and break-off distances differ, i.e. $d_{on}$/$d_{off}$ hysteretic mechanisms where $d_{on}$≠$d_{off}$ (12). The long range dissipative phenomena in (12) however can be further explored in terms of the $E_{dis}^*$ (Figure 7b) and $\Delta\Phi^*$ (Figure 7c) observables. First, note the characteristic step (increase) in the energy signal $E_{dis}^*$ at the distance $d=d_{on}$. This step coincides with a drop in $F_{ts}^*$ (Rec.) in Figure 7a when $d=d_{on}$. A similar step is also observed in the $\Delta\Phi^*$ signal at $d=d_{on}$ as shown in Figure 7c. These steps should be observed in experiments if long range hysteretic dissipative forces presenting a $d_{on}$/$d_{off}$ mechanism, such as that in (12), are present in the interaction. Note also the discontinuity at smaller distances (circles) in Figures 7b and 7c. These discontinuities coincide with the onset of perpetual formation of the capillary. Normalization in Figures 7b and 7c has been carried out with $E_{dis} \approx 27$ eV and $\Delta\Phi^* \approx 10°$. Finally, the energy dissipated at the $d_{on}<d<d_{off}$ range is recovered as a peak in $E_{dis}^*$ at $d=d_{on}$. In particular, from Figure 7b $E_{dis} \approx 2.5$ eV at $d=d_{on}$ in accordance with the dissipation due to (12).

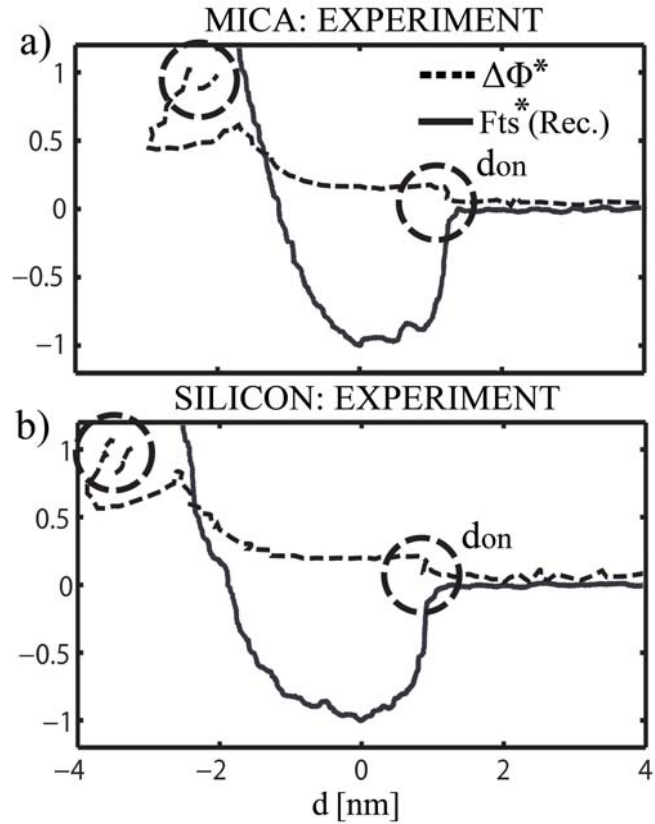

Figure 8. Experimental outcomes for a silicon tip interaction with a a) mica and a b) silicon sample. The recovered conservative force $F_{ts}^*$ (Rec.) is shown in continuous lines together with the $\Delta\Phi^*$ signals in dashed lines. A step-like increase occurs at a distance $d_{on}$ (circles) in the signal $\Delta\Phi^*$ and a corresponding drop in $F_{ts}^*$ (Rec.) for both samples. At smaller distances $d<0$ a second step in the $\Delta\Phi^*$ signal (circles) is observed as also predicted in Figure 7c. $A_0 \approx 31.5$ nm.

Experimentally, mica (Figure 8a) and silicon (Figure 8b) samples have been probed here to corroborate the appearance of step-like phenomena in the long range as that discussed in Figure 7 and modeled by (12). The samples have been probed at intermediate relative humidity RH≈40%. The force $F_{ts}^*$ (Rec.) and the corresponding $\Delta\Phi^*$ signals are shown with the use of continuous lines and dashed lines respectively. Normalization has been carried out with $F_{AD} \approx 1.7$ (mica) and $F_{AD} \approx 2.4$ (silicon) nN and $\Delta\Phi$ (mica) $\approx 10°$ and (silicon) $\Delta\Phi \approx 25°$ respectively. Several experimental outcomes are worth mentioning. First, note the steps for $d_{on}>0$ in the $\Delta\Phi^*$ signals (circles) for both samples as an abrupt drop in $F_{ts}^*$ (Rec.) also occurs. From the discussion of Figure 7, these abrupt variations or steps are identified with the onset of a long range dissipative force presenting a $d_{on}/d_{off}$ mechanism. Second, for both samples and by direct inspection of Figure 8 it is found that $d_{on}>0$ nm indicating that the steps, and associated dissipative processes, occur in the attractive regime where mechanical contact does not occur. Third, from the energy dissipation expression (2), the energies associated with the $d_{on}/d_{off}$ processs are $E_{dis}\approx 10$ eV and 40 eV for the mica and silicon samples respectively. Finally, experiments conducted at low humidity RH≈5% (data not shown) led to lower $d_{on}/d_{off}$ energies, i.e. a few eV, for the silicon tip-mica sample system while these energies did not show a significant decay with decreasing humidity for the silicon tip-silicon sample system. These results indicate that the dominant long range $d_{on}/d_{off}$ dissipative mechanism present in the silicon tip-mica sample interaction might originate from capillary interactions. This follows from the dependence of the associated energy dissipation $E_{dis}$ on humidity for the mica sample at $d \approx d_{on}$. Nevertheless, from the independence of the associated $E_{dis}$ on humidity for the silicon tip-silicon sample interaction, the main long range $d_{on}/d_{off}$ mechanism might originate from chemical affinity and the associated metastability of the approaching silicon surfaces. That is, atom and bond reorientation could be favorable on tip approach at $d \approx d_{on}$ and the rupture would be favorable on tip retraction (see scheme in Figure 1). The perpetual formation of neck and bond for the mica and silicon samples respectively is also found in the experiments in terms of a discontinuity in $\Delta\Phi^*$ for $d<0$ (see circles for $d<0$ and compare with Figure 7c).

## IV. Conclusion

A ratio in short range energy dissipation describing the interaction between a tip and a sample has been proposed and modeled with the use of simple dissipative force models[1, 4, 17]. This ratio involves dissipative processes that occur during mechanical contact and involving hysteretic and viscous forces. Through this ratio the relevant parameters controlling short range energy dissipation have been grouped according to whether they originate from dynamic, geometric or material properties. First, it has been shown via numerical simulations that it is possible to decouple the relative prevalence of short range hysteresis and short range viscosity via experimental observables. Then, by monitoring these experimental observables, it has been shown that the predictions of the ratio are in qualitative agreement with experiments on a mica sample. In particular, the results indicate that the functional relations described by the ratio in terms of experimental parameters, i.e. free and oscillation amplitude and tip radius, are consistent with nanoscale dissipation processes that are found experimentally.

Footprints of long range hysteretic dissipation have also been discussed via experimental observables. In particular, by varying the humidity it has been shown that it is possible to distinguish between long range hysteretic phenomena related to capillary interactions, i.e. silicon tip-mica sample, and that related to chemical affinity, i.e. silicon tip-silicon sample. Finally, it could be argued that the dissipative forces discussed analytically and in the simulations are simple and phenomenological in nature and might not describe the interaction exactly. Nevertheless, qualitative agreement between theory and experiment has been found throughout thus qualitatively confirming the functional relationships between velocity-related parameters, such as amplitude, and geometry-related parameters, such as tip radius. These findings and methodologies should assist in the robust formulation of nanoscale dissipative laws and bridge the gap between classical, atomistic and quantum theories.

Acknowledgements

We thank Maritsa Kissamitaki for producing the illustrations. This work was supported by the Ministerio de Economía y Competitividad (MEC), Spain, through the project MAT2012-38319-C02-01